\documentclass[fleqn,usenatbib]{mnras}
\usepackage{amsmath}
\usepackage{graphicx}
\usepackage{amssymb,txfonts}
\usepackage[T1]{fontenc}

\newcommand{\msun}{{\rm M}_{\sun}}

\newbox\grsign \setbox\grsign=\hbox{$>$} \newdimen\grdimen \grdimen=\ht\grsign
\newbox\simpropbox
\setbox\simpropbox=\hbox{\raise.5ex\hbox{$\propto$}\llap
   {\lower.5ex\hbox{$\sim$}}}\ht2=\grdimen\dp2=0pt
\def\simprop{\mathrel{\copy\simpropbox}}

\title[Jet model]{A simple analytical model of magnetic jets}

\author[A. A. Zdziarski et al.]
{Andrzej A. Zdziarski$^1$, {\L}ukasz Stawarz$^2$, Marek Sikora$^1$ and Krzysztof Nalewajko$^1$\\
$^1$Nicolaus Copernicus Astronomical Center, Polish Academy of Sciences, Bartycka 18, PL-00-716 Warszawa, Poland\\
$^2$Astronomical Observatory, Jagiellonian University, Orla 171, PL-30-244 Krak{\'o}w, Poland
}

\pagerange{\pageref{firstpage}--\pageref{lastpage}}
\pubyear{2022}

\begin{document}

\maketitle

\label{firstpage}

\begin{abstract}
We propose a simple analytical jet model of magnetic jets, in which radially-averaged profiles of main physical quantities are obtained based on conservation laws and some results of published GRMHD jet simulations. We take into account conversion of the magnetic energy flux to bulk acceleration in jets formed around rotating black holes assuming the mass continuity equation and constant jet power, which leads to the Bernoulli equation. For assumed profiles of the bulk Lorentz factor and the radius, this gives us the profile of the toroidal magnetic field component along the jet. We then consider the case where the poloidal field component is connected to a rotating black hole surrounded by an accretion disc. Our formalism then recovers the standard formula for the power extracted from a rotating black hole. We find that the poloidal field strength dominates over the toroidal one in the comoving frame up to large distances, which means that jets should be more stable to current-driven kink modes. The resulting magnetic field profiles can then be used to calculate the jet synchrotron emission.
\end{abstract}

\section{Introduction}
\label{intro}

Accreting black holes (BHs) of all sizes produce relativistic jets. The physical mechanisms producing the jets have been subject of intense research. The most promising mechanisms for the jet production appear to be those linked to magnetic fields \citep{BZ77,BP82}. These models also predict the jet power to scale with the BH mass (e.g., \citealt{HS03, Tchekhovskoy09, Tchekhovskoy10, Tchekhovskoy11}).

The observed power of extragalactic jets is often found to be very large, $\gtrsim \dot M_{\rm accr}c^2$, where $\dot M_{\rm accr}$ is the accretion rate (e.g., \citealt{Ghisellini14, Zamaninasab14, Pjanka17}). The only model capable to account for those powers appears to be that with electromagnetic extraction of the spin power \citep{BZ77} from a BH surrounded by a magnetically arrested disc (MAD; \citealt{BK74,Narayan03}), as demonstrated using GRMHD numerical simulations by \citet{Tchekhovskoy11} and \citet{McKinney12}. 

The structure of relativistic jets undergoing simultaneous acceleration and collimation has been investigated by complex semi-analytical relativistic MHD models, see \citet{Li92}, \citet{Vlahakis03}, \citet{Beskin06}, \citet{Lyubarsky09}. Numerical simulations with relativistic MHD have been performed by \citet{Komissarov07, Komissarov09}. Then, numerical GRMHD simulations have been done in a number of papers, see \citet{Tchekhovskoy15} and \citet{Davis20} for reviews. Here, we provide a simple analytical formalism reproducing main aspects of those models. We assume the continuity and energy conservation equations, and use the relationship between the toroidal and poloidal components of the magnetic field for the case of spin-power extraction based on the numerical simulations of \citet{Tchekhovskoy09}. However, we do not require the presence of a MAD.

Our model predicts the profiles of the radially-averaged toroidal and poloidal field components along the jet. We find the poloidal component to dominate over the toroidal one in the comoving frame up to large distances, which follows from the conversion of the toroidal field to the bulk acceleration taking place in the acceleration and collimation zone (ACZ). These profiles can then be used to calculate self-absorbed synchrotron emission from extended jets to model their flat radio spectra. Such models are generally based on the works of \citet{BK79} and \citet{Konigl81}, with a later formulation \citep{Zdziarski19b} of the synchrotron emission and self-absorption for arbitrary profiles of the bulk Lorentz factor, $\Gamma(z)$, the jet radius, $r(z)$, and the magnetic field strength, where $z$ is the distance from the BH centre.  

\section{The jet structure}
\label{structure}

The jet is launched from close vicinity of the BH and it first goes through the ACZ. Details of these processes are poorly understood. The jet radial profile depends sensitively on the profile of the external pressure, which is usually uncertain. Still, those profiles can be derived \citep[e.g.,][]{Vlahakis03}, but the resulting formalism is rather complex. Here, instead, we assume the profiles of both $\Gamma(z)$ and $r(z)$. In our model, $\Gamma$ increases until it reaches its terminal bulk Lorentz factor, $\Gamma_{\rm max}$, around which point the jet approaches a conical shape. We stress, however, that our general results below do not depend on these assumptions, and the profiles given below can be modified without affecting our formalism. Only a determination of the specific profiles of the magnetic field strengths requires the explicit specification of $\Gamma(z)$ and $r(z)$ in the ACZ, see Fig.\ \ref{B} below. We first assume,
\begin{equation}
\Gamma(z)=\begin{cases}a_\Gamma \left(z/r_{\rm H}\right)^{1/{q_1}}, & z_{\rm m}\equiv a_\Gamma^{-q_1}r_{\rm H}< z\leq z_{\rm t}\equiv (\Gamma_{\rm max}/ a_\Gamma)^{q_1} r_{\rm H};\cr
\Gamma_{\rm max}, & z > z_{\rm t},\cr\end{cases}
\label{Gamma}
\end{equation}
where $r_\mathrm{H} =[1+(1-a_*^2)^{1/2}]r_{\rm g}$ is the BH event horizon, $r_{\rm g}=G M/c^2$ is the gravitational radius, $a_*$ is the spin parameter, and $q_1\sim 2$ and $a_\Gamma< 1$ are constants. The jet starts to be accelerated at some distance, $z_{\rm m}$, above the stagnation surface. Its radius is assumed to be
\begin{align}
r(z)&=\begin{cases}a_r \left(z /r_{\rm H}\right)^{1/q_2}r_{\rm H}, & z_{\rm m}\leq z\leq z_{\rm t};\cr
\Theta_0 \left[z+(q_2-1)z_{\rm t}\right], & z > z_{\rm t},\cr\end{cases}
\label{radius}\\
\Theta_0&= \frac{a_r}{q_2} \left(\frac{a_\Gamma}{\Gamma_{\rm max}}\right)^{q_1(1-1/q_2)},\label{Theta}
\end{align}
where $a_r\sim 1$, $1<q_2<2$, and $\Theta_0$ is the asymptotic opening half-angle (assuming $\Theta_0\ll 1$; thus $\Theta_0\approx \tan\Theta_0$). The above formulae follow from the assumption that the transitions in both $\Gamma$ and $r$ take place at the same distance, $z_{\rm t}$, and from the continuity in both $r(z)$ and ${\rm d}r/{\rm d}z$ at $z_{\rm t}$. The latter, in particular, assures the continuity in the adiabatic loss rate. This differs from, e.g., the model of \citet{Ghisellini09}, in which the jet was assumed to have the radius $= \Theta_0 z$ in the conical part. We further require the product $\Theta_0\Gamma_{\rm max}$ to be given by a constant, as expected theoretically \citep{Tchekhovskoy09}. This is confirmed by \citet{Jorstad05}, \citet{Pushkarev09} and \citet{Clausen13}, who found the average values observed in extragalactic jets of $\Theta_0\Gamma_{\rm max}\approx 0.1$--0.2. The requirement of a constant $\Theta_0\Gamma_{\rm max}$ leads to
\begin{equation}
q_1=\frac{q_2}{q_2-1},\quad \Theta_0\Gamma_{\rm max}=\frac{a_r a_\Gamma}{q_2},
\label{TG}
\end{equation}
which we will assume hereafter. The former relationship is also given by \citet{Komissarov09}. On the other hand, their approximation for the radius has an additional dependence on the normalization on $q_2$ (denoted as $b$ in that paper). 

We assume the conservation of the rest mass-flow rate,
\begin{equation}
\dot M_{\rm j}=2 \upi r^2 \rho' c \beta \Gamma,
\label{mdot}
\end{equation}
where $\rho'$ is the mass density, and hereafter all primed quantities are in the jet frame. Thus, we assume the jet mass loading from the moment of its formation, which remains uncertain, see, e.g., \citet{ORiordan18}. We also assume the conservation of the jet power, $P_{\rm j}$; thus, we assume that the radiated power is $\ll\! P_{\rm j}$. The jet power equals the flux of the enthalpy, with the contributions from the ion rest energy, the internal enthalpy of both electrons and ions (consisting of their kinetic energy density, $u'$, and pressure, $p'$), and the Poynting flux integrated over the jet cross section,
\begin{equation}
P_{\rm j}= 2 \upi \rho' r^2 c^3 \beta \Gamma^2 + 2\upi(u'+p')r^2 c \beta \Gamma^2 + \frac{{B'_\phi}^2}{2}r^2 c \beta \Gamma^2,
\label{power}
\end{equation}
where the square of the toroidal field strength in the comoving frame, ${B'_\phi}^2$, is averaged over the jet cross section. Here, we included the power associated with the particle rest energy in the comoving frame. If most of the rest energy is in ions, it is supplied to the jet by the accretion flow from large distances. Then, the expression of the usable power would have the factor $\Gamma(\Gamma-1)$ instead of $\Gamma^2$ in the first term above\footnote{We note that \citet{Lucchini21} and \citet{Kantzas22} used a formula for the jet power in their {\tt bljet} model that is $\propto\Gamma^1$, which appears to be incorrect.}. If $\Gamma\gg 1$, this difference is negligible. 

The Poynting flux in an ideal-MHD jet (i.e., with the electric vector perpendicular to the magnetic one, ${\mathbf E}\cdot {\mathbf B}=0$) is $(1/4\upi){\mathbf B}\times ({\mathbf v}\times {\mathbf B})$. In a given point of the jet, this yields the energy-flow rate component parallel to the flow as $(1/4\upi)(\beta c B_\phi^2 - v_\phi B_{\rm p}B_\phi)$ where $B_\phi$ and $B_{\rm p}$ are the toroidal and poloidal field components in the local observer frame (hereafter the BH frame), respectively, and $v_\phi$ is the particle rotation velocity taken at an average jet radius. Hereafter, we neglect the effect of jet rotation, given it is slow beyond the Alfv\'{e}nic region, and the radiation contribution from that region is at most small, and thus we neglect the (negative) component of the jet power due to the poloidal field. Given that 
\begin{equation}
B_\phi=\Gamma B'_\phi,\quad B_{\rm p}= B'_{\rm p},
\label{transform}
\end{equation}
we obtain equation (\ref{power}). Then, the jet magnetisation parameter, defined as the ratio of the Poynting flux to the rest-energy flux, is
\begin{equation}
\sigma \equiv \frac{{B'_\phi}^2}{4 \upi \rho' c^2}.
\label{sigma}
\end{equation}
By dividing the expression for the jet power by that for the rest-mass flow, we obtain the Bernoulli equation, expressing conversion of the magnetic energy into the bulk motion. Hereafter, we assume the contribution from the internal enthalpy of the particles, i.e., the second term in the right-hand side of equation (\ref{power}), to be small. This is often found to be satisfied in modelling individual jet sources, see, e.g., \citet{Ghisellini14}, \citet{Zdziarski22c}. This also justifies the neglect of the internal energy density term in the denominator of equation (\ref{sigma}). Thus, the expression for the magnetisation parameter following from the Bernoulli equation becomes
\begin{equation}
\sigma(z) =\frac{\Gamma_{\rm max} \left(1+\sigma_{\rm min}\right)}{\Gamma(z)}-1,
\label{bernoulli}
\end{equation}
where a decreasing $\sigma$ corresponds to a conversion of the Poynting flux into the jet acceleration, and $\sigma_{\rm min}$ is achieved at $\Gamma=\Gamma_{\rm max}$. The jet power is then,
\begin{equation}
P_{\rm j}=\dot M_{\rm j}c^2 \left(1+\sigma_{\rm min}\right)\Gamma_{\rm max}.
\label{power2}
\end{equation}
Note that this includes the $\dot M_{\rm j}c^2$ term supplied by the accretion flow. We then use equations (\ref{mdot}), (\ref{sigma}) and (\ref{bernoulli}) to calculate the (square-averaged) toroidal magnetic field strength along the jet,
\begin{equation}
{B'_\phi}^2(z)=\frac{2 \dot M_{\rm j}c\sigma(z)}{r(z)^2 \beta(z) \Gamma(z)}.
\label{Bphi}
\end{equation}
Note that since the jet is accelerated at the expense of the enthalpy flux of the toroidal component of the magnetic field, ${B'_\phi}$ decreases in the ACZ faster than $\propto r(z)^{-1}$, by $(\sigma/\Gamma)^{1/2}$. We note that the formulation of equations (\ref{mdot}--\ref{Bphi}) is quite general. The value of $B'_\phi$ follows uniquely from the assumed profiles of $\Gamma(z)$ and $r(z)$ and the assumption that the acceleration is at the expense of the magnetic energy, and it is independent of the details of this process. An additional assumption is that the internal energy is small compared to the rest energy, but in most actual models it is at most comparable.

The minimum value of the jet magnetisation parameter, $\sigma_{\rm min}$, is achieved in the conical part of the jet. We consider here two assumptions about this parameter,
\begin{equation}
\sigma_{\rm min}=
\begin{cases}(\Theta_0 \Gamma_{\rm max}/s)^2=\left[a_r a_\Gamma/(s q_2) \right]^2, & {\rm case\ I};\cr
{\rm free}, & {\rm case\ II},\cr\end{cases}\
\label{sigma_min}
\end{equation}
Case I assumes the validity of the approximate relation between the magnetisation, Lorentz factor and opening angle obtained by \citet{Tchekhovskoy09} and \citet{Komissarov09}, where $s$ is constrained by causality to $\lesssim 1$. In this case, we can take $\Theta_0\Gamma_{\rm max}$ to be a free parameter instead of $\sigma_{\rm min}$, which (for a given $s$) implies the value of $\sigma_{\rm min}$ (and $a_r a_\Gamma$). However, case I is satisfied only if the jet acceleration and collimation are dominated by ideal MHD processes. While this is likely for $\sigma\gtrsim 1$, conversion of the Poynting flux to the kinetic energy leading to $\sigma \ll 1$ can be driven by dissipative processes (see, e.g., the discussion in \citealt{ZSPT15} and \citealt{Pjanka17}). Therefore, we also consider the case of a free $\sigma_{\rm min}$.  

In order to find the evolution of the jet poloidal magnetic field, $B'_{\rm p}$, along the outflow, we use the conservation of the jet magnetic flux, $\Phi_B \equiv 2\upi \int_0^{r(z)} {\rm d} r' B'_{\rm p}(r')$, where $r'$ denotes the running radius of the jet cross section, $r'\leq r(z)$. Then, the relationship between $B'_\phi$ and $B'_{\rm p}$ depends on the angular velocity of the field lines, $\Omega_{\rm f}$. One possibility is the jet formation by magnetic fields attached to the accretion disc \citep{BP82}, in which case $\Omega_{\rm f}$ is related to the disc rotation. See, e.g., \citet{Sadowski10} for an analysis of the required inclination of the field lines. Here, we consider magnetic fields attached to a rotating BH (a split monopole magnetosphere), resulting in energy extraction from the BH rotation \citep{BZ77}. In this case, $\Omega_{\rm f}=c a_*\ell/2 r_{\rm H}$, where $\ell\sim 1/2$ is the ratio of $\Omega_{\rm f}$ to the BH angular frequency, $\Omega_{\rm H}$. Then,
\begin{equation}
B'_{\rm p}(r')=\frac{c\beta\Gamma}{\Omega_{\rm f} r'}B'_\phi(r'),
\label{bp_phi}
\end{equation}
see, e.g., equation (22) of \citet{Tchekhovskoy09}, where the ideal MHD is assumed. In this case, the conversion of the Poynting flux to the bulk acceleration causes a differential redistribution of poloidal field lines. As a result, both poloidal and toroidal components of the magnetic field are not uniform along the jet radius. Their radial dependencies obtained in the numerical simulations of \citet{Tchekhovskoy09} can be approximated as, cf.\ equations (17--20) of \citet{ZSPT15},
\begin{equation}
\frac{B'_{\rm p}(r')}{B'_{\rm p}(z)}\approx \frac{\sigma}{1+\sigma} \left[\frac{r'}{r(z)}\right]^{\frac{-2}{1+\sigma}}\!\!\!, \quad \frac{B'_\phi(r')}{B'_\phi(z)}\approx  \left(\frac{2\sigma}{1+\sigma}\right)^{\frac{1}{2}} \left[\frac{r'}{r(z)}\right]^{\frac{\sigma-1}{1+\sigma}},
\label{radial}
\end{equation}
where $B'_{\rm p}(z)=\langle {B'_{\rm p}(r')}\rangle\equiv \Phi_B/(\upi r^2)$, and  ${B'_\phi}(z)=\langle [B'_\phi(r')]^2\rangle^{1/2}$ (as noted above). This implies, using equations (\ref{bernoulli}--\ref{Bphi}) and (\ref{bp_phi}--\ref{radial}),
\begin{equation}
B'_{\rm p}(z) \approx B'_\phi(z) \frac{2^\frac{3}{2} r_{\rm H}\beta\Gamma}{a_* \ell r}\left(\frac{1+\sigma}{\sigma}\right)^\frac{1}{2}
=\frac{4r_{\rm H} \left(\beta P_{\rm j}/c \right)^\frac{1}{2}}{a_* \ell r^2}
\label{Bp}
\end{equation}
In order to calculate the total field strength in the comoving frame accurately, we would need to add in quadrature the toroidal and poloidal contributions of equation (\ref{radial}) and then average over the cross section. Here, we simply approximate it as ${B'}^2 \approx {B'_{\rm p}}^2+{B'_\phi}^2$. 

Assuming the validity of equations (\ref{Gamma}--\ref{TG}) in the conical part, we have
\begin{equation}
\frac{B'_{\rm p}(z)}{ B'_\phi(z)}\approx \frac{2^{3/2} r_{\rm H}\Gamma_{\rm max}^2 q_2}{a_* \ell a_r a_\Gamma\sigma_{\rm min}^{1/2} z},\quad z_{\rm eq}\approx \frac{2^{3/2} r_{\rm H}\Gamma_{\rm max}^2 q_2}{a_* \ell a_r a_\Gamma \sigma_{\rm min}^{1/2}},
\label{Bratio}
\end{equation}
where we assumed $\sigma_{\rm min}<1$. The two components become equal at $z_{\rm eq}$. We see that $z_{\rm eq}\gg r_{\rm H}$ generally. In the case of $q_2=2$, $z_{\rm eq}/z_{\rm t}\approx 2^{5/2}a_\Gamma /(a_*\ell a_r \sigma_{\rm min}^{1/2})$. This can be either greater or less than unity depending on the values of $\sigma_{\rm min}$, $a_\Gamma/a_r$ and $a_*$. We note that \citet{Vlahakis03} and \citet{Komissarov09} found $z_{\rm eq}/z_{\rm t}<1$. However, we point out that our definition of $z_{\rm t}$ is specific to the assumption about the $\Gamma(z)$ dependence given by equation (\ref{Gamma}).

\begin{figure*}
\centerline{ \includegraphics[width=6cm]{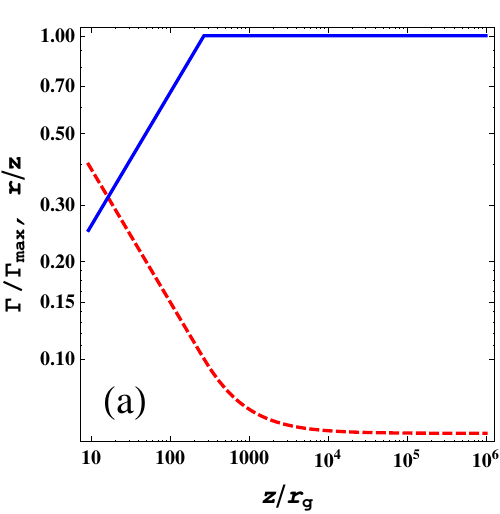} \includegraphics[width=6cm]{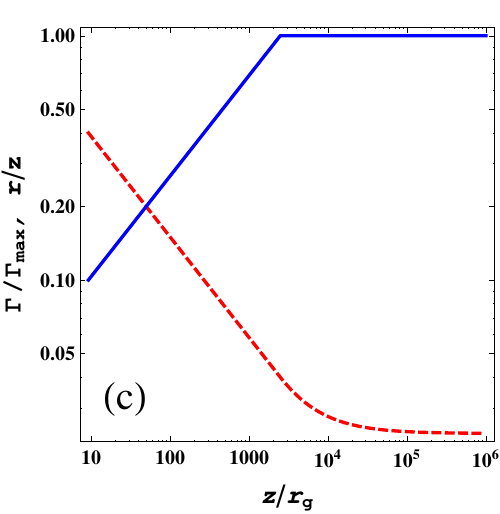} \includegraphics[width=6cm]{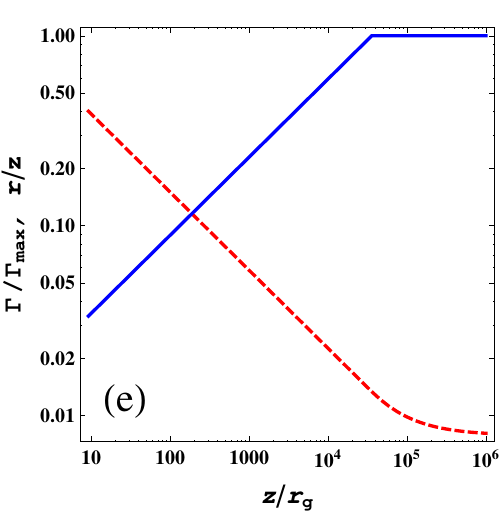}} 
\centerline{ \includegraphics[width=6cm]{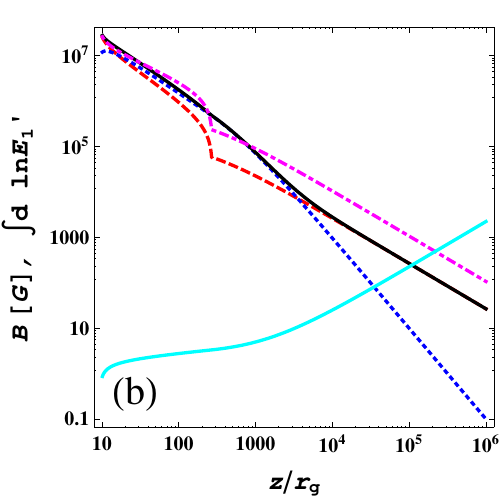} \includegraphics[width=6cm]{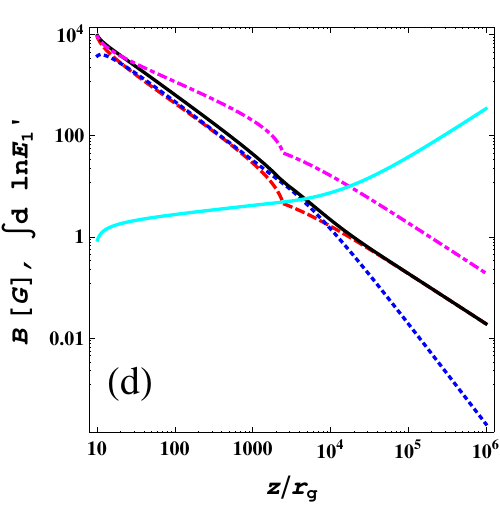} \includegraphics[width=6cm]{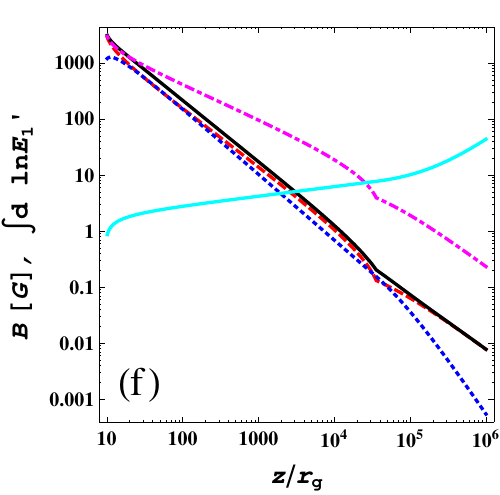}} 
\caption{Examples of the dependencies on the distance of (top panels) $\Gamma/\Gamma_{\rm max}$ and $r/z=\tan\Theta$, and (bottom panels) the magnetic field components, for $P_{\rm j}=0.1 L_{\rm E}$, $\ell=1/2$, $a_*= a_r= 1$, $a_\Gamma=0.4$, $q_2=1.7$ (implying $\Gamma_{\rm max}\Theta_0\approx 0.24$, $z_{\rm m}\approx 9.3 r_{\rm g}$), $q_1=q_2/(q_2-1)$. (a, b) $M=10\msun$, $\Gamma_{\rm max}=4$, $\sigma_{\rm min}=10^{-2}$, $\Theta_0\approx 3.4\degr$, (c, d) $M=10^8\msun$, $\Gamma_{\rm max}=10$, $\sigma_{\rm min}=(\Gamma_{\rm max}\Theta_0)^2$, $\Theta_0\approx 1.4\degr$, and (e, f) $M=10^9\msun$, $\Gamma_{\rm max}=30$, $\sigma_{\rm min}=0.1$, $\Theta_0\approx 0.45\degr$. The profiles of $\Gamma/\Gamma_{\rm max}$ and $r/z$ are shown by the blue solid curves and the red dashed curves, respectively, and the profiles of the poloidal, comoving-frame toroidal and total magnetic field strengths are shown by the blue dotted, red dashed and black solid curve, respectively. The profile of the BH frame toroidal field strength is shown by the magenta dot-dashed curve. All components of $B$ scale $\propto(P_{\rm j}/M)^{1/2} M^{-1/2}$. The jet approaches the conical shape above $z_{\rm t}=\Gamma_{\rm max}^{q_1} r_{\rm H}$. In the bottom panels, we see the kinks in the dependencies of the toroidal field, which are artefacts of the assumed $\Gamma(z)$ profile. The solid cyan curves in the bottom panels show the growing amplitudes of linear comoving-frame kink perturbation calculated from equation (\ref{growth}).
} \label{B}
\end{figure*}

Alternatively, we could have used the fact that the average poloidal and toroidal magnetic field components are expected to be close to equipartition (in the local observer frame) at the Alfv\'{e}n surface, located at some distance $z_{\rm A}$ from the centre, meaning $B_{\rm p}(z_{\rm A})=B_\phi(z_{\rm A})$. Then, we could assume that at the Alfv\'{e}n surface the jet is strongly magnetised, $\sigma(z_{\rm A})\gg 1$. Under this condition, the Alfv\'{e}n surface approaches the light cylinder \citep{Appl93,Lyubarsky10}. This approach yields the scaling similar to, but more complex than, that of equation (\ref{Bp}), and it yields somewhat lower values of $B'_{\rm p}$. 

The conserved poloidal magnetic flux, $\Phi_B= \upi B'_{\rm p}r^2$, is then,
\begin{equation}
\Phi_B= \frac{4\upi r_{\rm H} \left[(1+\sigma_{\rm min})\Gamma_{\rm max}\dot M_{\rm j} c\right]^{1/2}}{a_* \ell}=\frac{4\upi r_{\rm H} \left(P_{\rm j}/c \right)^{1/2}}{a_* \ell},
\label{mag_flux}
\end{equation}
where we set $\beta=1$ consistent with our assumption of the negligible rest energy flow contribution to the jet power, requiring $\Gamma\gg 1$, equation (\ref{power}). The power of spin extraction from the BH is $\propto \Omega_{\rm f}(\Omega_{\rm H}-\Omega_{\rm f})\Phi_B^2\propto \ell(1-\ell)\Omega_{\rm H}^2 \Phi_B^2$ \citep{BZ77}, which is maximized at $\ell=1/2$, which is usually assumed. Here, we give this power following equation (3.6) of \citet{Tchekhovskoy15}, but expressing it for an arbitrary value of $\ell$,
\begin{equation}
P_{\rm BZ}\approx \frac{\ell(1-\ell) c}{24\upi^2 r_{\rm H}^2}\Phi_B^2 a_*^2
=\frac{2\ell(1-\ell)}{3\ell^2}P_{\rm j},
\label{PBZ}
\end{equation}
where the first equality is accurate for $a_*\lesssim 0.95$ \citep{Tchekhovskoy10}. For $a_*\gtrsim 0.95$, the multiplicative correction factor is given by equation (3.8) of \citet{Tchekhovskoy15}, see also \citet{Camilloni22}. In the second equality, we used equation (\ref{mag_flux}). Noting that $P_{\rm j}=P_{\rm BZ}+\dot M_{\rm j}c^2$ and for relativistic jets $P_{\rm BZ}\gg \dot M_{\rm j}c^2$, we find that equation (\ref{PBZ}) is satisfied for $\ell=\sqrt{2/5}\approx 0.6$, which is close to 1/2. This shows the overall self-consistency of our determination of the poloidal flux. The value of $\ell\neq 1/2$ appears to be due to both the numerical coefficient in equation (\ref{mag_flux}) being approximate, and the numerical coefficient in the first equality in equation (\ref{PBZ}) depending on the geometry of the magnetic field \citep{Tchekhovskoy15}.   

In order to relate $P_{\rm j}$ and $\Phi_B$ to the accretion rate, $\dot M_{\rm accr}$, we introduce the standard definition of a dimensionless magnetic flux, $\phi_{\rm BH}$,
\begin{equation}
\phi_{\rm BH}\equiv \Phi_{\rm BH}/(\dot M_{\rm accr}r_{\rm g}^2 c)^{1/2},
\label{phibh}
\end{equation}
where $\Phi_{\rm BH}$ is the magnetic flux threading the BH on one hemisphere
(e.g., \citealt{Tchekhovskoy11,McKinney12,Davis20}). The magnetic flux conservation implies that $\Phi_{\rm BH}=\Phi_B$. Published GRMHD simulations yield the maximum value of $\phi_{\rm max}\approx 70(1-0.38 a_* r_{\rm g}/r_{\rm H})h_{0.3}^{1/2}$, where $h_{0.3}$ is defined by the half-thickness of the disc being $h_{\rm disc}=r_{\rm disc} 0.3 h_{\rm 0.3}$ (e.g., \citealt{Davis20}). The maximum $\phi_{\rm BH}$ is achieved for a spinning BH surrounded by a MAD. With equation (\ref{PBZ}) at $\ell=1/2$, we have
\begin{equation}
P_{\rm j}\approx 1.7\left(\frac{\phi_{\rm BH} r_{\rm g}}{40 r_{\rm H}}\right)^2 a_*^2 \dot M_{\rm accr} c^2,
\label{Pphi}
\end{equation}
where we have normalized it to the value of $\phi_{\rm max}$ at $a_*=1$. The numerical coefficient is slightly larger compared to equation (13) in \citet{Davis20}, where it is 1.3. This difference appears to result from the numerical coefficients in our equations (\ref{Bp}) and (\ref{PBZ}) being approximate. We stress, however, that all of the formulae in our paper do not require the presence of a MAD. In our formulation, the jet power, $P_{\rm j}$, is a free parameter, and thus the dimensionless magnetic flux can assume any value $\leq \phi_{\rm max}(\dot M_{\rm accr}r_{\rm g}^2 c)^{1/2}$. Then, $\dot M_{\rm accr}$ can be estimated from the accretion efficiency, $\epsilon_{\rm accr}$, and the accretion luminosity, $L_{\rm accr}$, as $L_{\rm acc}\equiv \epsilon_{\rm accr}\dot M c^2$. We can also express $L_{\rm accr}$ in terms of the Eddington luminosity, $L_{\rm E}=4\upi G M m_{\rm p} c/\sigma_{\rm T}$, which is given here for pure hydrogen, and where $m_{\rm p}$ is the proton mass and $\sigma_{\rm T}$ is the Thomson cross section. We can also define the jet production efficiency, $\epsilon_{\rm j}$, by $P_{\rm j}\equiv \epsilon_{\rm j}\dot M_{\rm accr} c^2$. 

Fig.\ \ref{B} shows some examples of the dependencies on the distance along the jet of the bulk Lorentz factor, $\Gamma$, and the jet opening angle, $\Theta$, which we show as $r/z=\tan\Theta$, shown in the top panels, and the magnetic field strengths, shown in the  bottom panels. Here, we assume the validity of equations (\ref{Gamma}--\ref{TG}). We assume $a_*= a_r=1$, $a_\Gamma=0.4$, $q_2=1.7$. We see that $B'_{\rm p}$ dominates over (or is similar to) $B'_\phi$ up to relatively large distances. As noted after equation (\ref{Bphi}), this follows from the fast decrease of $B'_\phi\propto (\sigma/\Gamma)^{1/2}r^{-1}$ in the ACZ. We have then $B'_\phi\simprop z^{-(1/q_1+1/q_2)}$, $B_\phi\simprop z^{-1/q_2}$, $B'_{\rm p}\propto z^{-2/q_2}$ in the ACZ, and $B'_\phi\propto B_\phi\propto z^{-1}$, $B'_{\rm p}\propto z^{-2}$ in the conical part. All components of $B$ are $\propto(P_{\rm j}/M)^{1/2} M^{-1/2}$. The kink in the dependencies for $B_\phi$ and $B'_\phi$ is artefact of the assumption that the jet stops its acceleration sharply at $z_{\rm t}$, which results in the derivative of $\Gamma(z)$ being not continuous. Similar plots are shown in \citet{Vlahakis03} and \citet{Komissarov09} for parameters relevant to gamma-ray bursts. As we noted above, they found $B'_{\rm p}=B'_{\phi}$ to take place in the ACZ, which difference is related to our specific assumptions of equations (\ref{Gamma}--\ref{TG}) as well as to their assumed $\sigma_{\rm min}\approx 1$. We also note that equation (\ref{TG}) gives the opening angle of $\Theta_0=a_r a_\Gamma/(q_2 \Gamma_{\rm max})$, which can be large for a low $\Gamma_{\rm max}$. This is then in contrast to the observations of accreting BHs in stellar binary systems, where low values of both $\Gamma_{\rm max}$ and $\Theta_0$ are commonly observed \citep{Miller-Jones06, Zdziarski22a}. This discrepancy can be resolved assuming a low enough $a_\Gamma$. 

We then compare our obtained value of the poloidal magnetic field strength in the jet close to the horizon to that in a MAD disc, $B_z$. We follow the estimate of the MAD magnetospheric radius of equation (1) of \citet{Xie19}, 
\begin{equation}
B_z^2\approx \frac{r_{\rm g}\dot M_{\rm accr}c^2}{r_{\rm disc}^3 v_r h_{\rm disc}/r_{\rm disc}}\approx \frac{r_{\rm g}P_{\rm j}}{r_{\rm disc}^3 v_r a_*^2 0.3 h_{0.3}^2},
\label{bmad}
\end{equation}
where in the second equality we used the jet power of equation (\ref{Pphi}) at the maximum flux and neglecting the numerical coefficient of 1.7, and $v_r$ is the accretion radial velocity. We note that $v_r$ will not be equal that for a usual accretion disc, given the `choked' nature of MAD accretion. We then compare this estimate at $r_{\rm disk}=r =  r_{\rm H}$ to the $B'_{\rm p}$ of equation (\ref{Bp}). We find that the equality is obtained for 
\begin{equation}
(v_r/c)h_{0.3}^2\approx (\ell/5)(r_{\rm H}/r_{\rm g}),
\label{equalB}
\end{equation}
which requires $h_{0.3}^2 v_r/c\sim 0.1$. A similar estimate was done in \citet{Tchekhovskoy15}. 

\section{Summary and discussion}
\label{discussion}

We presented a simple analytical model of magnetized jets. We parametrized the ACZ by profiles of the bulk Lorentz factor and radius [$\Gamma(z)$ and $r(z)$, equations (\ref{Gamma}) and (\ref{radius}), respectively], and assumed the jet is conical and at a constant speed beyond the ACZ. However, our general results do not depend on the specific form of $\Gamma(z)$ and $r(z)$. We then assumed a constant mass flow rate through the jet, equation (\ref{mdot}), which is equivalent to assuming the jet is mass-loaded at its base. Next, we assumed the jet power is conserved, which corresponds to the radiated power being $\ll P_{\rm j}$, which is often satisfied. Given the above assumptions, we obtained the profiles of the radius-averaged magnetization parameter and the toroidal component of the magnetic field strength, with an additional assumption of the internal particle density being less than the rest energy density. We then related the poloidal component of the magnetic field to the toroidal one assuming the rotation of the magnetic field lines related to the BH angular velocity, and took into account the radial profiles of the field components being not uniform. This gave us the conserved magnetic flux. This flux, when used in the formula for the BH spin-extraction power, yields then the initially assumed jet power, showing the self-consistency of our approach. The main free parameters of our model are $P_{\rm j}$, $\Gamma_{\rm max}$ and $\sigma_{\rm min}$. 

Using the above formulation, we found that the poloidal component of the magnetic field can dominate up to one order of magnitude beyond the length, $z_{\rm t}$, of the ACZ, see equation (\ref{Bratio}) and Fig.\ \ref{B}. The cause of this effect is the conversion of the toroidal magnetic field component into the bulk acceleration in the ACZ, which results in $B'_\phi\propto r^{-1} (\sigma/\Gamma)^{1/2}$, which yields in turn the decline with the radius a significantly faster than $r^{-1}$. This effect was already noticed in \citet{Vlahakis03} and \citet{Komissarov09}. Furthermore, we stress that the detailed profiles depend on the profiles of $\Gamma(z)$ and $r(z)$, and the $B'_{\rm p}$--$B'_{\phi}$ equipartition may occur at lower values of $z$ than those shown in Fig.\ \ref{B}. Still, we note the overall dependence on the assumed $\Gamma(z)$ and $r(z)$ is relatively weak. At the base of the jet, we have $\Gamma\sim 1$ and $r\sim r_{\rm H}$ while in the conical part we have $\Gamma_{\rm max}\gg 1$ and $\sigma_{\rm min}< 1$. The profiles of $B'_\phi$ and $B'_{\rm p}$ have to then smoothly join the two regimes. 

We have made several assumptions, listed above. While some of them correspond to the standard magnetic jet model, e.g., \citet{Tchekhovskoy09}, the presented formalism provides a framework in which these assumptions can be modified. E.g., we can modify equation (\ref{mdot}) to allow for mass loading along the jet, or modify equation (\ref{power}) to account for radiative losses. We can also modify the angular velocity of the field lines to that of the model of the jet powered by the magnetic fields attached to the accretion disc \citep{BP82}. Furthermore, we could assume the geometric profile of $r(z)$ and could calculate the profile of $\Gamma(z)$ using the momentum conservation along a representative streamline (e.g., assuming self-similarity). However, this would involve divergence of tensors, and can be done only for individual stream lines requiring the knowledge of the values of $B'_\phi$ and $B'_{\rm p}$ along them. This would significantly complicate the presented formalism (cf.\ \citealt{Vlahakis03}). 

We then consider the jet stability for the $m = 1$ kink modes driven by poloidal electric current, see \citet{Appl00}. In the force-free case with the magnetic pitch, $\mathcal{P}' \equiv r' B'_{\rm p}(r') / B'_\phi (r')$, independent of $r'$ (as assured by equation \ref{bp_phi}), the growth time scale for a linear perturbation is $\tau_{\rm kink}' \simeq 7.5 \mathcal{P}' / v_{\rm A,p}'$, where $v_{\rm A,p}' = c \sqrt{ \sigma_{\rm p} / (1+\sigma_{\rm p}) }$ is the comoving-frame Alfv\'en speed at the jet axis and $\sigma_{\rm p} = {B'_{\rm p}}^2(r'=0) / (4\upi \rho' c^2)$. With equation (\ref{bp_phi}), $\mathcal{P}' \approx 2\beta\Gamma r_{\rm H}/(a_*\ell)$, and then 
\begin{equation}
\tau'_{\rm kink}(z) \approx \frac{15\beta\Gamma r_{\rm H}}{a_*\ell v_{\rm A,p}'}.
\label{kink}
\end{equation}
We can then calculate the growth of the perturbation amplitude, $E'_1$, 
of the electric field along the jet,
\begin{equation}
\int {\rm d}\ln E'_1=\int_{z_{\rm m}}^z {\rm d}z'/[\Gamma(z')\beta(z')c \tau_{\rm kink}'(z')].
\label{growth}
\end{equation}
We note that $\tau_{\rm kink}'$ depends on the value of $B_{\rm p}'(r'=0)$, which may be divergent according to the profiles adopted in equation (\ref{radial}). In such case, we approximate $v_{\rm A,p}' \approx c$, although we note that in reality the profile of $B_{\rm p}'(r')$ should approach a finite value at $r'=0$, which would lower $v_{\rm A,p}'$ and increase $\tau_{\rm kink}'$. The numerical solutions to equation (\ref{growth}), which we include in the bottom panels of Fig.\ \ref{B} correspond to the fastest growth of perturbation amplitude allowed by the functions $\mathcal{P}'(z)$ and $\Gamma(z)$. We find that the perturbation amplitude shows two phases -- an initial phase of slow growth followed by a phase of fast growth. The transition between these two phases corresponds roughly to the distance at which $B_\phi'$ starts to dominate over $B_{\rm p}'$, which is usually somewhat beyond $z_{\rm t}$.

In M87, \citet{Asada12} and \citet{Asada14} found $q_2\approx 1.7$ and $q_1\approx q_2/(q_2-1)\approx 2.4$, respectively, and $z_{\rm t}\approx 2\times 10^5 r_{\rm g}$. However, the initial acceleration was found to be in the non-relativistic regime, and the above power-law dependence of $\Gamma$ was reproduced only for $z\gtrsim 2\times 10^4 r_{\rm g}$. Then, \citet{Hada13} found $q_2\approx 1.8$ at $z\gtrsim 200 r_{\rm g}$, in an approximate agreement with \citet{Asada12}, but they found a steeper profile with $q_2\approx 1.3$ at lower $z$. These results point out to further complexity of the jet shape and velocity profile in observed sources. That, however, affects only our equations (\ref{Gamma}--\ref{Theta}) and the relation between $z_{\rm t}$ and $\Gamma_{\rm max}$, which could be accounted for in a future work. 

\section*{ACKNOWLEDGEMENTS}
We thank Arieh K{\"o}nigl, Serguei Komissarov and Sasha Veledina for valuable discussions, and the referee for valuable suggestions. We acknowledge support from the Polish National Science Center under the grants 2019/35/B/ST9/03944 and 2021/41/B/ST9/04306. 

\section*{Data Availability}
There are no new data associated with this article.
\bibliographystyle{mnras}
\bibliography{../allbib} 

\label{lastpage}
\end{document}